\documentstyle[12pt]{article}

\begin{document}
\begin{titlepage}
\begin{center}
\today
\hfill    LBL-35683 \\
\hfill    UCB-PTH-94/12 \\
\hfill    ACT-20/94 \\
\vskip .1 in
{\large \bf THE ALL-GENUS STRING EFFECTIVE ACTION}
\footnote{This work was supported in part by the Director, Office of
Energy Research, Office of High Energy and Nuclear Physics, Division of
High Energy Physics of the U.S. Department of Energy under Contract
DE-AC03-76SF00098,  in part by the National Science Foundation under
grant PHY-90-21139 and in part by ICSC-World Laboratory.}
\vskip .2 in
{\bf Rulin Xiu} \\
\vskip .1 in
{\em Theoretical Physics Group\\
    Lawrence Berkeley Laboratory\\
      University of California\\
    Berkeley, California 94720\\
\vskip 12pt
Astroparticle Physics Group \\
Houston Advanced Research Center \\
The Woodlands, Texas 77381, USA \\

}
\end{center}

\vskip .1in

\begin{abstract}
 We use the off-shell string effective action method developed by
 E.S. Fradkin and A.A. Tseytlin
 to obtain the formula for all-genus string effective action
with and without
compactification at the low-energy approximation
in the massless background fields.
We find that for the bosonic string, one can determine
the dilaton vacuum expectation value from the all-genus effective action
because of the nontrivial dependence of potential energy on dilaton.
For compactified four-dimensional string models,
if one requires that the target-space dilaton field lie on a
 K\"ahler manifold, we obtain a constraint which will
specify the worldsheet dilaton in terms of the constant background fields.
We also show that under this constraint, the tree-level
k\"ahlar potential and superpotential are not changed by the higher-genus
effect. This proves again the non-renormalization theorem
for a string moving in massless background fields in the low-energy
approximation.
\end{abstract}
\end{titlepage}
\renewcommand{\thepage}{\roman{page}}
\setcounter{page}{2}
\mbox{ }

\vskip 1in

\begin{center}
{\bf Disclaimer}
\end{center}

\vskip .2in

\begin{scriptsize}
\begin{quotation}
This document was prepared as an account of work sponsored by the United
States Government. While this document is believed to contain correct
 information, neither the United States Government nor any agency
thereof, nor The Regents of the University of California, nor any of their
employees, makes any warranty, express or implied, or assumes any legal
liability or responsibility for the accuracy, completeness, or usefulness
of any information, apparatus, product, or process disclosed, or represents
that its use would not infringe privately owned rights.  Reference herein
to any specific commercial products process, or service by its trade name,
trademark, manufacturer, or otherwise, does not necessarily constitute or
imply its endorsement, recommendation, or favoring by the United States
Government or any agency thereof, or The Regents of the University of
California.  The views and opinions of authors expressed herein do not
necessarily state or reflect those of the United States Government or any
agency thereof of The Regents of the University of California and shall
not be used for advertising or product endorsement purposes.
\end{quotation}
\end{scriptsize}

\vskip 2in

\begin{center}
\begin{small}
{\it Lawrence Berkeley Laboratory is an equal opportunity employer.}
\end{small}
\end{center}

\newpage
\renewcommand{\thepage}{\arabic{page}}
\setcounter{page}{1}

In this paper, we use the  off-shell string effective action
method developed by E.S. Fradkin
and A.A. Tseytlin \cite{off}
to derive the form of all-genus effective action in the
background fields with and without compactification at the low-energy
 approximation. We will first review their derivation of the bosonic string
effective action using this method. Then
we will show  that this method can be easily generalized to all
genus case and we will derive the form of all-genus effective action for
bosonic string, superstring and heterotic string with or
without compactification.
We find that  the
dependence of  coupling constants
 on dilaton remains unchanged with a redefinition of dilaton,
 but the kinetic energy
of dilaton is modified. For a bosonic string, from all-genus
effective action one can determine the vacuum expectation value (VEV) of
world-sheet
dilaton field. For four-dimensional string models, we find that
the requirement of  target-space dilaton field on a K\"ahler manifold
constrains the worldsheet dilaton to take a specific value
determined by the constant string background fields.
We also find
that with a redefinition of dilaton
fields,
the all-genus K\"ahler potential and superpotential
 remain the same as the tree-level ones.
 This gives another proof of
the non-renormalization theorem for effective string theory
with massless background fields in the low-energy approximation.

One of the big goals of string phenomenology is to determine the low energy
effective
theory from string dynamics. To achieve this,  one needs to
study string theories in a curved background (\i.e. interactive string
theories). So far there are three methods of studying the interactive
string theory, namely the S-matrix method \cite{onshell,smat}, the
beta function method
\cite{call} and the string effective action method \cite{off}. The
S-matrix approach is to construct an effective field theory that
yields the same scattering amplitudes as those given by the full
string theory at the low-energy limit.
In the beta function method, one  derives the background field equations
of motion and the
effective lagrangian from the conformal invariance of string theory.
The off-shell string effective action method, developed by E.S. Fradkin and
A.A. Tseytlin, directly deduces the effective action from the path integral
formulation of interactive string theory.

The starting point of E.S. Fradkin and A.A. Tseytlin's method
  is the covariant effective action
\[\Gamma[C,G_{ij},B_{ij}...],\]
which is defined through  the string path integral
to be:
\begin{eqnarray}
\Gamma[C,G_{ij},B_{ij}...] & \equiv & \sum_{\chi}
e^{\sigma\chi}\int_{M_{\chi}^2} {\cal D}g_{\alpha \beta }{\cal D}X^i e^{-I}.
\end{eqnarray}
Here the path integral is over all maps $X(z)$ of the worldsheet
$M_{\chi}^2$ into the spacetime manifold $M^D$, and over all
two-dimensional metrics $g_{\alpha \beta }$ on the worldsheet, and $\chi
\,=\,(4\pi)^{-1}\int d^2z \sqrt{g} R \,=\, 2-2n$ is the Euler number
of $M_{\chi}^2$.  The constant $\sigma$ is related to the
dimensionless string theory coupling constant $ g_c\,=\, e^{-\sigma}$.
Note that $\sigma$ can be absorbed into a constant part of dilaton field C
 and is fixed
when the vacuum value of dilaton field C is determined by the full theory.
So the worldsheet dilaton field C can also be thought of as the string coupling
constant.
  $I$ is  the string action including
the coupling of string to massless string excitations  (or for a string
moving in the manifold with the background fields).
For the closed bosonic string, in which the dilaton
$C$, graviton $G_{ij}$ and antisymmetric tensor $B_{ij}$ form the
massless level of string spectrum, the string action including these
``excited" modes is:
\begin{eqnarray}
I & = &  \int d^2\sigma [  (4\pi)^{-1}\sqrt{g}
R\,C(X) \nonumber \\ &+ &
(4\pi \alpha' )^{-1}\sqrt{g}\,g_{\alpha \beta}\partial_{\alpha}
X^i\partial_{\beta }X^j\,G_{ij}(X)
+ i\epsilon^{\alpha \beta }\partial_{\alpha }
X^i\partial_{\beta }X^j\,B_{ij}(X) ].
\end{eqnarray}
The effective action $\Gamma$ contains all the information about
quantum string theory. It is a generating functional for the
correlation functions of the background fields: $ C, G_{ij},
B_{ij}$,\ldots .  From the target space point of view, $\Gamma$ is the
quantum field theory effective action of fields $ C, G_{ij}, B_{ij},
...$ .  So the above string path integral yields a direct derivation
of the effective field theory action and the effective lagrangian from
string dynamics.

To compute the effective action defined above,
one first  extracts the integral over a ``center of mass" collective
coordinate by splitting $X^i\,=\,x^i+\eta^i, \,\,\mbox{where } x^i=\,$constant,
\[ \int\,[dX]\,\,=\,\,\int\,d^Dx\,\int[d\eta]. \]
Then one calculates the ``effective
action'' for a generalized $\sigma$-model (with the ``internal'' space $M^D$)
defined on a curved world-sheet,
\[ exp(-W[g,G,B])=\int [d\eta ]\,
exp(-I[x+\eta,G(x+\eta),B(x+\eta),C(x+\eta),g]). \]
And finally one averages over all possible metrics on the world-sheet and sums
over all possible topologies. Considerations based on covariance and
the Weyl anomaly in an interacting theory determine the general
structure of $W[g,C,G,B]$:
\begin{eqnarray}
W & = & \epsilon^{-1}\beta\,\int\,d^2x\,\sqrt{g}R\,+\gamma\,\int\,
d^2zd^2z'(R\sqrt{g})_z\partial^{-2}_{zz'}(R\sqrt{g})_{z'}, \nonumber\\
\beta & = & 4a_1, \nonumber \\
\gamma & = & a_1+
a_2\alpha'({\cal R}(x)-\frac{1}{12}H_{ijk}H^{ijk}+
4\partial_iC\partial^iC) \nonumber \\ & & +
a_3\alpha'^2[{\cal R}_{...}(x)+...]^2+... ,
\end{eqnarray}
where $\epsilon\,=\,d-2\rightarrow 0$, R is the worldsheet curvature scalar
and $\cal{R}$ is the target-space curvature scalar. From the one- and two-loop
calculations, one finds:
\begin{eqnarray}
 a_1 & = & D/96\pi,\nonumber \\
\gamma & = & D/96\pi\,-\,
(\alpha'/64\pi)({\cal R}-\frac{1}{12}H_{ijk}H^{ijk}+
4\partial_iC\partial^iC)\nonumber \\ & &\,+\,O(\alpha'^2{\cal R}^2).
\end{eqnarray}

Next one integrates $e^{-W}$ over the metrics on a closed surface $M^2_{\chi}$
under the coordinate gauge:
\begin{equation}  g_{\mu \nu}\,=\,e^{2\rho}\hat{g}_{\mu \nu},
\end{equation}
here $\hat{g}_{\mu \nu}$ has a
constant curvature.  One finds that the logarithm of the ghost determinant has
the same structure as the above W, the total $\gamma$ is:
\begin{equation}
\gamma\,=\,(96\pi)^{-1}
[ D - 26 - \frac{3}{2}\alpha'({\cal R}-\frac{1}{12}H_{ijk}H^{ijk}+
4\partial_iC\partial^iC) \,+\,0(\alpha'^2{\cal R}^2) ]. \label{eq:gamma}
\end{equation}
The derivation so far depends only on local geometric properties, and the
result is true for arbitrary genus world-sheet.
At tree level ($\chi$=2)
we have:
\begin{eqnarray}
\Gamma^{tree}& \sim & \int d^Dx e^{-2C}
\int [d\rho] \nonumber \\ & & \times \exp\left(-\gamma
\int d^2zd^2z'\times (\hat{R}\sqrt{\hat{g}}-2\hat{\partial}^{2}\rho)
_z\hat{\partial}^{-2}_{zz'}(\hat{R}\sqrt{\hat{g}}-2\hat{\partial}^{2}
\rho)_{z'}\right).
\end{eqnarray}
The integration over $[\rho]$ under the fixed surface area gauge condition
yields a trivial  factor which does not depend on $\gamma$ , we get:
\begin{equation}
\Gamma ^{tree} \sim
 \int d^Dx e^{-2C}\,exp(-\gamma\int d^2zd^2z'(\hat{R}\sqrt{\hat{g}})_z
\hat{\partial}^{-2}_{zz'}(\hat{R}\sqrt{\hat{g}}_{z'}).
\end{equation}
For a sphere, $\int \hat{R}\hat{\partial}^{-2}\hat{R}=16\pi$. For a critical
string in the  low energy
approximation, \i.e., $\alpha'$ small, one gets:
\begin{eqnarray}
\Gamma[G,C,B] & = & -M^D\int d^Dx\sqrt{G} e^{-2S}\nonumber \\ & &
\times  [1+\frac{1}{4}\alpha'({\cal
R}+4\partial_iC\partial^iC-\frac{1}{12}H_{ijk}H^{ijk})+...].
\end{eqnarray}
Here $M\sim (\alpha')^{1/2}$ is a normalization mass. Making the Weyl
rescaling $G_{ij}\rightarrow G_{ij} \exp[4C/(D-2)]$, we obtain:
\begin{eqnarray}
\Gamma[G,C,B] &=&\int d^Dx\sqrt{G}\{-(4/k^2\alpha) exp[4C/(D-2)]
+ \nonumber \\
& & k^{-2} [-{\cal
R}+(4/(D-2))(\partial_iC)^2 -\frac{1}{12}H_{ijk}^2exp[-8C/(d-2)]]\},
\nonumber \\
k\, & \sim & \, (\alpha')^{(D-2)/4}.
\end{eqnarray}
This result agrees with that deduced from the S-matrix method and the
beta-function method. E.S. Fradkin and A.A. Tseytlin also extend the above
string effective action approach to superstring theories. In this paper, we
will assume the above method can be applied to obtain
the effective action for supersymmetric and heterotic strings.

In the following, we will generalize the above 0-genus result to an all-genus
one.
This is possible because
the derivation of (\ref{eq:gamma})
depends only on  local geometric properties and
it is true  for a world-sheet with arbitrary topology. So we can easily
extend the above tree-level result perturbatively  to  an
all-genus result. For example,  for an n-genus world-sheet,
the string effective action is:
\begin{eqnarray}
\Gamma^{n-genus}&\sim & \int d^Dx e^{-2(1-n)C}
\int_{M_{\chi}} dm_j \int [d\rho] \nonumber \\
& &
exp\left(-\gamma\int d^2zd^2z'\times
(\hat{R}\sqrt{\hat{g}}-2\hat{\partial}^2\rho)_z\hat{\partial}^{-2}_{zz'}
(\hat{R}\sqrt{\hat{g}}-2\hat{\partial}^2\rho)_{z'}\right)
\nonumber \\
&\sim &  \int d^Dx e^{-2(1-n)C}  \int_{M_{\chi}} dm_j
\nonumber \\ & &
exp(-\gamma \int d^2z d^2z'(\hat{R}\sqrt {\hat{g}})
_z\hat{\partial}^{-2}_{zz'}(\hat{R}\sqrt{\hat{g}})_{z'}).
\end{eqnarray}
Here  $m_j$ are the moduli parameterizing the n-genus worldsheet.
In the low-energy approximation,
 these expressions can be rewritten in a form similar
to the tree-level
results:
\begin{eqnarray}
\Gamma^{n-genus}[G,B,C,...]
&=& \int d^Dx\sqrt{G} e^{(2n-2)C}\big\{-(4/k^2\alpha) b_n
\nonumber \\ & & + k^{-2}d_n[-{\cal
R}-4(\partial_iC)^2
 + \frac{1}{12}H_{ijk}^2] +\ldots \big\},
\end{eqnarray}
with $b_n$ and $d_n$  defined to be:
\begin{eqnarray}
 b_n&=& \int _{M_{\chi}} dm_j,  \\
 d_n&=& -(64\pi)^{-1}\int _{M_{\chi}} dm_j
\int d^2zd^2z'
(\hat{R}\sqrt{\hat{g}})_z\hat{\partial}^{-2}_{zz'}
(\hat{R}\sqrt{\hat{g}})_{z'}.
\end{eqnarray}
We see $b_n$ and $d_n$
 are some constants which depend only on the topology of a worldsheet.
Summing over all-genus worldsheets, we get:
\begin{eqnarray}
\Gamma[G,B,C]&=&\int d^Dx\sqrt{G}\big\{-(4/k^2\alpha)
\sum_{n=0,1,...}e^{-(2n-2)C}b_n \;\;\;\; \nonumber \\ & &
+k^{-2}[-{\cal R}-4(\partial_iC)^2
-\frac{1}{12}H_{ijk}H^{ijk}]\sum_{n=0,1,...}d_n\,e^{-(2n-2)C}\big\}.\;\;\;\;
\end{eqnarray}
With the Weyl rescaling $G_{ij}\rightarrow G_{ij}h^{\frac{-2}{D-2}}(C)$,
\[h(C)=\sum_{n=0,1,...}d_n\,e^{-(2n-2)C},\] and
\[f(C)=\sum_{n=0,1,...}b_n\,e^{-(2n-2)C},\]
 we obtain:
\begin{eqnarray}
\Gamma[G,B,C]=\int d^Dx\sqrt{G}\big\{-(4/k^2\alpha)
f(C)h^{-\frac{D}{D-2}} + \nonumber \\
 k^{-2}[-{\cal R}-(4-\frac{D-1}{D-2}h'^2(C))(\partial_iC)^2
- \frac{1}{12}h^{-\frac{4}{D-2}}H_{ijk}H^{ijk}] \big\}.
\end{eqnarray}
We see that the dilaton field is modified  in the all-genus case. We express
the effective action in terms of the modified dilaton $C\rightarrow h$:
\begin{eqnarray}
\Gamma[G,B,C] = \int d^Dx\sqrt{G}\big\{-(4/k^2\alpha)
f(C)h^{-\frac{D}{D-2}}(C) + \nonumber \\
 k^{-2}[-{\cal R}-(\frac{4}{h'^2(C)}-\frac{D-1}{D-2})(\partial_ih)^2
- \frac{1}{12}h^{-\frac{8}{D-2}}H_{ijk}H^{ijk}] \big\},
\end{eqnarray}
with
\[ h'(C)=\frac{\partial h(C)}{\partial C}. \]
We see that the high-genus effect modifies the
tree-level kinetic
energy of dilaton. But the coupling constants,
 for example the coupling constant in
front of
$H_{ijk}$ and, if gauge fields exist, the gauge coupling constant
have the same dilaton dependence under the redefined dilaton field.
 Another interesting result from the all-genus effective action
is  that for bosonic string
 because of the nontrivial dependence of vacuum energy on the dilaton
field,
one can
determine the  worldsheet dilaton (VEV) from:
\begin{equation}
\frac{\partial}{\partial C}[f(C)h^{-\frac{D}{D-2}}(C)]=0
\end{equation}

Next we generalize the tree-level result to an all-genus one for the
supersymmetric string or heterotic string.
In these cases, the above derivation still applies
except now $b_n=0, n=0,1,...$. This is  because
the integration of supermoduli over a constant
yields zero. So
the all-genus effective action  for the supersymmetric strings  is of the form:
\begin{eqnarray}
\Gamma [C,G_{ij},B_{ij}] & \equiv & \int d^Dx {\cal L}
= \int d^Dx\, { \cal L}^{tree} ( C, G_{ij}, B_{ij}, A_i... )h(C),
\nonumber \\
 & = & \int d^Dx\sqrt{G}\big\{
 k^{-2}[-{\cal R}-(\frac{4}{h'^2(C)}-\frac{D-1}{D-2})(\partial_i h)^2
\nonumber \\ & &
+ \frac{1}{12}h^{-\frac{4}{D-2}}H_{ijk}H^{ijk}]
+\frac{1}{4}h^{-\frac{2}{D-2}}F_{ij}F^{ij} \big\}.
\end{eqnarray}
Here again, the coupling constants remain unchanged under the redefined
dilaton but the dilaton kinetic energy is modified.
Notice for superstring, the determination of dilaton VEV is not possible.

Now we proceed to calculate
the  all-genus effective action with some dimensions
of spacetime compactified. In this case,
  the  coordinate $X^I$ on compactified space decomposes into three parts:
\( X^I = x^I+\eta^I+X_c^I \), i.e. the center of mass part $x^I$, the
excitation part
$\eta^I$ and the zero modes from compactification $X_c^I$. So now we get:
\begin{equation}
\int {\cal D}X^I\,=\,\int {\cal D}x^I\int {\cal D}\eta^I\int {\cal
D}X^I_{c}.
\end{equation}
Here  $\int {\cal D}X^I_{c}$ is the summation over the zero-modes
coming from compactification.
 The effective action becomes:
\begin{eqnarray}
\Gamma[C,G_{ij},B_{ij}] =
\sum_{\chi=2,0,-2,...}e^{-\sigma \chi} \int d^Dx\,
\int {\cal D}'X\,{\cal D}g\,e^{-I}\, \int {\cal D}X^I_{c}\, e^{-A_{0}}
\end{eqnarray}
Here we define $A_0$ to be the action of the zero-modes coming from
compactification.  Notice that the oscillation mode contribution is
the same for the compactified and uncompactified coordinates.  So the
effective action with compactification will be the same as the one we
derived before, except for one extra factor coming from the summation
of compactification zero-modes. For n-genus worldsheet, we define this
factor to be:
\begin{equation}
Z_n(G_{IJ},B_{IJ},A_I,m_j)\,=\, \int_{M_{\chi}} {\cal D}X^I_{c}\,
e^{-A^n_{0}},
\end{equation}
with  $A_0^n$ representing the n-genus worldsheet action of the
zero-modes from compactification and the $m_j$ are the moduli
parameterizing the n-genus world-sheet. For torus or orbifold compactification,
$Z_n(G_{IJ},B_{IJ},A_I,m_j)$ can be
 calculated.  For example, for torus compactification,  at the tree level,
$Z_0(G_{IJ},B_{IJ},A_I,m_j)$ is trivial:
\(Z_0(G_{IJ},B_{IJ},A_I,m_j)\,\,=\,\,1/Vol(\Lambda)\), where $Vol(\Lambda)$ is
the volume of the compactified space.  At the one-genus level, this
has been calculated for heterotic string \cite{torus} to be:
\begin{eqnarray}
Z_1(G_{IJ},B_{IJ},A_I,\tau, \bar{\tau}) & = &\int_{\chi=0} {\cal
D}X^I_{c}\, e^{-A_{0}^{1-loop}} \nonumber \\
& = & \sum_{\scriptstyle L \in \Lambda
\scriptstyle P \in \Lambda^{*}} \sum_{ W \in \frac{{\rm spin}(32)}{Z_2}}
\bar{q}^{p_L^2/2} q^{p_R^2/2}.
\end{eqnarray}
Here $q=e^{2i\pi\tau}$, $\tau$ is the modulus parameterizing the
worldsheet torus and
\begin{eqnarray}
p^I_{\scriptstyle L,\scriptstyle R} &=&
\frac{1}{2}G^{IJ}P_J \mp L^i + G^{IK}B_{KL}L^L+
\frac{1}{2}G^{IJ}A_J^aW^a
+\frac{1}{4}G^{IJ}A_I^aA_J^aL^K,\\ p^I_L &=&p^I+A_K^I
n^K.\end{eqnarray}
 $G_{IJ}, B_{IJ}$ and $A_I$ are the constant string background fields
that parameterize string vacua.
We see that for high-genus string worldsheet,  $Z_n(G_{IJ},B_{IJ},A_I,\tau,
\bar{\tau})$ has a nontrivial dependence on these constant background
fields.  With the all-genus result derived above,  the
formula for effective action with compactification is:
\begin{eqnarray}
\Gamma[C,G,B,A] & = &
\sum_{\chi=2,0,-2,...}e^{-\sigma \chi}
\int d^Dx\,\int {\cal D}'X\,{\cal D}g\,e^{-I}\,
Z_n(G_{IJ},B_{IJ},A_I,m_j) \nonumber \\
& = & \int d^Dx\,
{ \cal L}_{10}^{\mbox{tree}} ( C, G,
B)h(C,G_{IJ},B_{IJ},A_I^{\alpha}).
\end{eqnarray}
Now the function $h$ depends on the worldsheet dilaton C and other
constant string background fields $G_{IJ}, B_{IJ},$ and $A_I$. It is defined
to be:
\begin{equation}
h(C,G_{IJ},B_{IJ},A_I^{\alpha})\,
=\sum_{n=0,1,...}d_n\,e^{(2-2n)C},
\end{equation}
with
\begin{eqnarray}
d_n = -\frac{1}{64\pi}\int _{{\cal M}_h} dm_j
Z_n(G_{IJ},B_{IJ},A_I;m_j) \int d^2zd^2z'
(\hat{R}\sqrt{\hat{g}})_z\hat{\partial}^{-2}_{zz'}
(\hat{R}\sqrt{\hat{g}})_{z'}.
\end{eqnarray}

The four-dimensional lagrangian ${\cal L}$ is specified by three functions:
 K\"ahler potential $K$, superpotential $W$ and  $f$ function.
To derive these functions, here we use
the dimensional reduction method introduced by
E. Witten \cite{wdr}. We get:
\begin{eqnarray}
\Gamma[C,G_{ij},B_{ij}, A_i]=
\int d^4x\, \big\{\frac{1}{2}R^{(4)}-3\partial_{\mu}\sigma\partial^{\mu}
\sigma+2\partial_{\mu}C\partial^{\mu}C
\nonumber \\
-\frac{9}{16h^2}
\partial_{\mu}h\partial^{\mu}h-\frac{3}{2}e^{-2\sigma}h^{\frac{1}{2}}
e^{-2\sigma}(\partial_{\mu}a-\frac{1}{\sqrt 2}ik\phi_xD_{\mu}\phi^x)^2
\nonumber \\
-\frac{3}{4}h^{\frac{1}{2}}e^{6\sigma}H^{\mu\nu\rho}H_{\mu\nu\rho}
-\frac{1}{4}h^{\frac{1}{4}}e^{3\sigma}TrF_{\mu\nu}F^{\mu\nu}
-3e^{-\sigma}h^{\frac{1}{4}}D_{\mu}\bar{\phi_x}D^{\mu}C^x
\nonumber \\
-\frac{8}{3}g^2h^{\frac{1}{4}}e^{-5\sigma}|\frac{\partial W}
{\partial \phi_x}|^2-\frac{9}{2}(g^2/f)h^{\frac{1}{4}}e^{-5\sigma}
\sum_i(\bar{\phi},\lambda^i \phi)^2
-8k^2g^2h^{\frac{1}{2}}e^{-6\sigma}|W|^2\big \}.
\end{eqnarray}
Here $\sigma$ is a scalar field related to ``breathing mode'',
$\phi_x$ are some charged scalar fields in a string model.
The axion ``D'' is defined by:
\[ h^{\frac{1}{4}}e^{3\sigma}H_{\mu\nu\rho}=\epsilon_{\mu\nu\rho\sigma}
\partial^{\sigma}D.\]
 $W$ is the cubic ``superpotential'':
\[ w=8\sqrt{2}d_{xyz}\phi^x\phi^y\phi^z.\]
Identifying the gauge coupling constant with  dilaton field, we find that
  dilaton and moduli in four-dimension, S and T, should be defined
by:
\begin{eqnarray}
S=e^{3\sigma}h^{\frac{1}{4}}+3i\sqrt D, \nonumber \\
T=e^{\sigma}h^{-\frac{1}{4}}-i\sqrt{2}a+\bar{\phi}_x\phi^x.
\end{eqnarray}
To have a K\"ahler geometry, the coefficients in front of the
$\partial_{\mu}D\partial^{\mu}D$ term and
 $\partial_{\mu}S_r\partial^{\mu}S_r$ term
 (here $S_r$ is the real part of dilaton)
should be the same. This requires that:
\begin{equation}
2\partial_{\mu}C\partial^{\mu}C-
\frac{9}{16h^2}\partial_{\mu}h\partial^{\mu}h=
-\frac{1}{16h^2}\partial_{\mu}\partial^{\mu}h
\end{equation}
or:
\begin{equation}
[\frac{\partial h(C,G_{IJ},B_{IJ},A_I)}{h\partial C}]^2=4.
\end{equation}
It is interesting that the requirement of a K\"ahler manifold for the
four-dimensional target-space dilaton field
leads to the determination of the worldsheet dilaton C in terms of
the constant string background fields in a string model.
But the breathing mode $\sigma$ still remains undetermined.

It is easy to see that, under the above constraint, the all-genus
superpotential and K\"ahler potential remain the same as the tree-level
ones. This can be viewed as another proof of string non-renormalization
theorem \cite{dine,Mak,SU(3)}. Notice
that from our calculation the all-genus $f$ function,
which is the gauge coupling function,
also remains the same as the tree-level one. But this result does not
contradict the result of Kaplunovsky~\cite{kap}.
This is because in our derivation we
take into account only massless background fields, while the
string threshold correction comes from the integration over massive string
modes. This is a limitation of our result.

\vskip 28pt
\noindent {\bf Acknowledgement}
\vskip 12pt

I would like to thank Mary K. Gaillard and Dimitri Nanopoulos
for their great inspiration. This work benefits a lot from
discussing with many people, to name just a few here,
Stanley Mandelstam, Orlando Alvarez and J. Cohn. I am very grateful
that Paul Watts helped me understand some of mathematics involved
in this paper and  Mary K. Gaillard pointed out that it is necessary to do Weyl
rescaling and taught me how to do it.

This work was supported in part the Director, office of Energy Research,
Office of Energy Research, Office of High Energy and Nuclear Physics,
Division of High Energy Physics
of the U.S. Department of Energy under Contract DE-AC03-76SF00098, in part
by the National Science Foundation under PHY-90-21139 and in part by
ICSC-World
Laboratory.

\end{document}